\documentclass[aip,reprint,twocolumn,showpacs,floatfix,preprintnumbers,superscriptaddress,amsmath,amssymb]{revtex4-2}
\usepackage{amsmath}
\usepackage{amssymb}
\usepackage{amsfonts}
\usepackage{graphicx}
\usepackage{epsfig}
\usepackage{epstopdf}
\usepackage{dcolumn}
\usepackage{bm}
\usepackage{array}
\usepackage{mathtools}
\usepackage{url}
\usepackage{subfig}
\usepackage{lipsum}
\usepackage{ragged2e}
\usepackage[colorlinks,bookmarks=false,citecolor=blue,linkcolor=blue,urlcolor=blue]{hyperref}

\begin{document}

\title{Comparative Analysis of Phase Noise for different configurations of Bragg lattice for an Atomic Gravimeter with Bose-Einstein Condensate}
\author{Pranab Dutta}
\affiliation{Department of Physics, Indian Institute of Science Education and Research, Pune 411008, Maharashtra, India}
\author{S. Sagar Maurya}
\affiliation{Department of Physics, Indian Institute of Science Education and Research, Pune 411008, Maharashtra, India}
\author{Korak Biswas}
\affiliation{Department of Physics, Indian Institute of Science Education and Research, Pune 411008, Maharashtra, India}
\author{Kushal Patel}
\affiliation{Department of Physics, Indian Institute of Science Education and Research, Pune 411008, Maharashtra, India}
\author{Umakant D. Rapol}
\email {Electronic mail: umakant.rapol@iiserpune.ac.in}
\affiliation{Department of Physics, Indian Institute of Science Education and Research, Pune 411008, Maharashtra, India}

\begin{abstract}
{

We perform a comparative study of the phase noise induced in the lasers used for Bragg diffraction in a Bose-Einstein condensate based quantum gravimeter where the Bragg beams are generated using two different configurations. In one of the configurations, the Bragg beams that form the moving optical lattice are generated using two different acousto-optic modulators. In the second configuration, the Bragg beams are generated using a single acousto-optic modulator carrying two phase locked frequencies. The second configuration shows a suppression of phase noise by a factor of 4.7 times in the frequency band upto 10 kHz, the primary source of noise, which is the background acoustic noise picked up by optical components and the optical table. We report a sensitivity of $\mathrm{99.7}$ $\mathrm{\mu Gal/\sqrt{Hz}}$ for an interferometric time of 10 ms.}

\end{abstract}

\maketitle
\section*{Introduction} 

\addcontentsline{toc}{section}{Introduction} 
Atom Interferometers (AI) have shown to be a promising tool in precision measurement over an Optical Interferometer(OI). Theoretically, the atom interferometer’s sensitivity is about ten orders of magnitude larger than the optical interferometer \cite{PhysRevA.48.3186} due to the rest-mass of the interfering atoms compared to photons. In the last three decades, the source of atoms for an AI moved on from a beam of hot atoms to a cloud of cold atoms \cite{PhysRevLett.66.2689,PhysRevLett.78.2046,PhysRevLett.66.2693}for increased sensitivity. At present, AI using  cold atoms has been successfully demonstrated for utilization in gravimetry \cite{PhysRevA.88.043610,peters1999measurement,le2008limits}, magnetometry \cite{PhysRevA.82.061602,smith2011three}, rotation sensing \cite{PhysRevLett.116.183003, PhysRevLett.114.063002}, inertial navigation \cite{10.1117/12.2228351, geiger2011detecting}, measurement of fundamental constants \cite{PhysRevLett.106.080801, rosi2014precision} and tests of general relativity \cite{PhysRevLett.98.111102, rosi2017quantum}. Over the last two decades, significant advancements have been made in the practical application of these AI based quantum sensors \cite{bongs2019taking, alonso2022cold,dutta2023decade}

AI, based on cold atoms is limited by the coherence length of atoms leading to degradation of fringe contrast in the interference signal \cite{szigeti2012momentum} which finally decreases the sensitivity of the system. AI based on Bose-Einstien Condensate (BEC) have been shown to lead to higher  contrast owing to the larger brightness and coherence length of BECs. The BEC's high number density is a concern for precision measurement as it introduces interaction-induced dephasing due to mean-field energy \cite{PhysRevLett.100.080405}. This can be avoided  by providing sufficient time of flight before the interferometric pulse sequence is initiated, whereby the mean field interaction becomes negligible \cite{PhysRevLett.82.4569}. For the measurement of absolute gravity, BEC based atom interferometers have been shown to be as accurate as conventional cold atom based interferometers. Atom chip based BEC AI even can  potentially reduce the electronic and optical complexity \cite{PhysRevLett.117.138501} and have shown accuracies below sub$-\mu$Gal (1Gal = $\mathrm{10^{-2}m/{s^2}}$) \cite{PhysRevLett.117.138501}. There have been continuous efforts worldwide to reduce its size, weight and power  for transportability on ground as well as for deployment in space. 

The majority of AI systems based on Bose-Einstein condensates (BECs) necessitate intricate optical configurations for system realization. These configurations introduce phase noise, which can be mitigated through the implementation of passive and active isolation systems \cite {PhysRevLett.117.138501}, or via the utilization of optical phase-locking loop techniques (OPLL) \cite{10.1063/1.5001963}.  
Here, we report the implementation of two configurations for generating Bragg beams on a BEC based AI to measure the gravity using $\mathrm{^{87}Rb}$ \cite{PhysRevLett.117.138501, PhysRevLett.117.203003, PhysRevA.84.033610} and made a comparative analysis of the phase noise reduction. Conventionally two separate AOMs are used to realize the Bragg beams required for AI \cite{10.1063/1.5001963}. In this work we achieved a significant reduction of phase noise using a single AOM \cite{PhysRevLett.119.263601} which reduces the optical complexity in comparison to the conventional method. By utilizing this approach, the Bragg beams and its associated pulses, carrying both frequencies, shared the same optical path, including mirrors, and optics. As a result, it mitigated the phase noise by significantly reducing the common mode. This  reduction in common-mode phase noise proved to be highly effective, enabling us to extend the interferometric time to more than 10 ms, a task that was challenging with the two distinct AOM configurations in our setup. Furthermore, we characterized this reduction of noise by carrying out detailed study of the phase noise in both the configurations \cite{10.1063/1.5001963}.

\begin{figure}[ht]
	\centering
	\includegraphics[width=\linewidth]{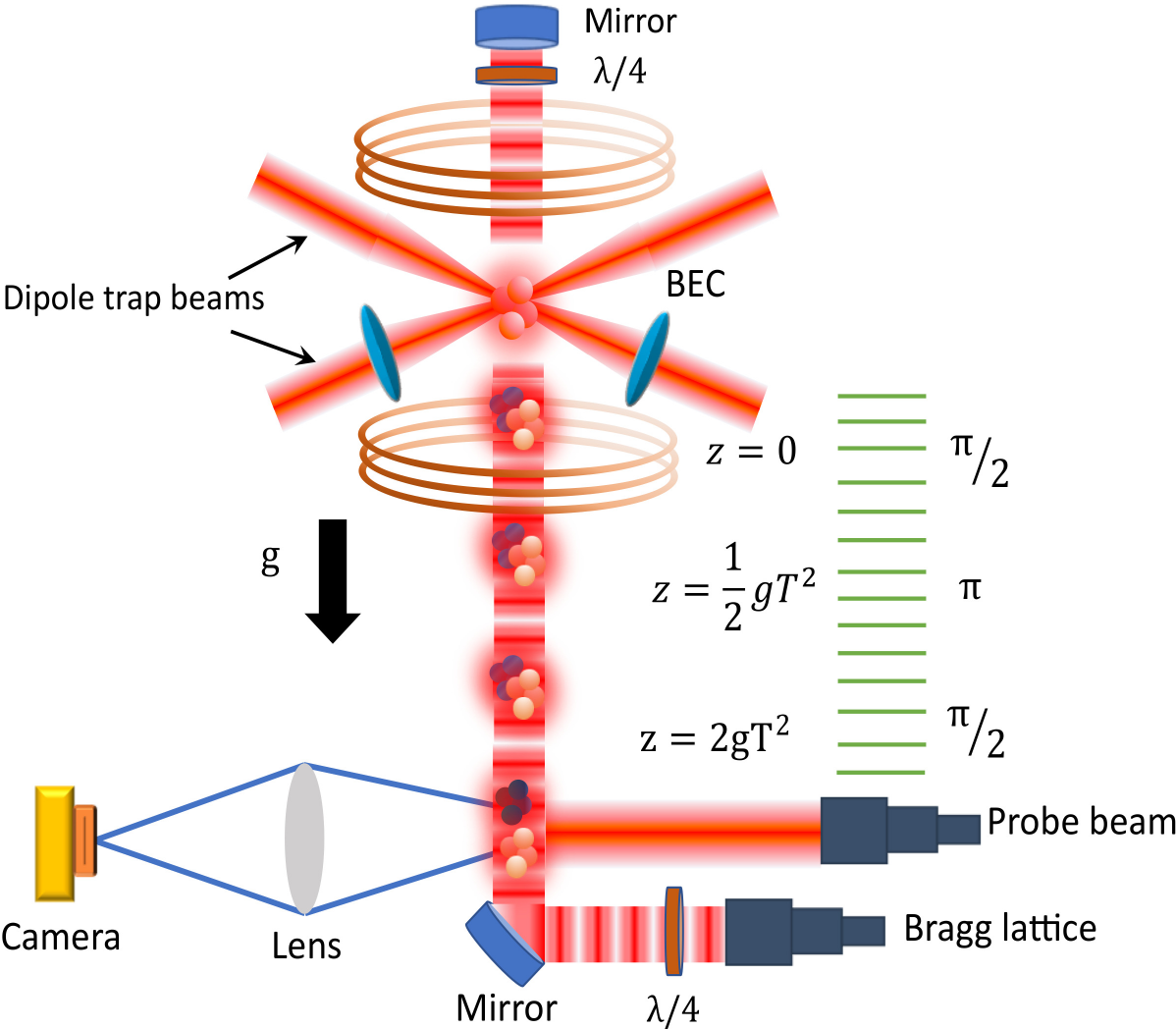} 
	\captionsetup{%
		justification=centerlast,
		labelfont =bf,
		singlelinecheck=false
	}
	\caption [] {Schematic diagram of BEC based atom interferometer for gravimetry and its space time trajectories.The diagram represent the formation of BEC at the center of the magneto-optical trap which is realized in a hybrid trap using a dipole trap and the magnetic field. The BEC is allowed to free fall in the absence of the trap potential and allowed to evolve under the interferometric pulses. The green dashed lines represent the three consecutive interferometric pulses for the realization of Mach-Zehnder interferometer with BEC. Figure adapted from our previous work\cite{dutta2023decade}. }
	\label{fig1}
\end{figure}
\section{Quantum Gravimeter}

\subsection{Measurement Principle}
The operation of an atom interferometer relies on spatial manipulation of the atomic wavepacket to achieve the interferometric fringe signal. This manipulation is usually achieved using laser light arranged in different configurations. Some of the commonly used techniques are Bragg diffraction, Bloch oscillations and Raman diffraction \cite{Sarkar22,PhysRevA.85.013639}. In an atom interferometer the atomic wavepacket is to be split and recombined using optical 'beam splitters' and 'optical mirrors' that change the momentum of the atomic wavepacket. In our gravimeter, we utilize the technique of Bragg diffraction, which is extensively described in references  \cite{PhysRevA.61.041602,altin2013precision}. This technique involves coupling two momentum states, namely  $\mathrm{p_0}$ and $\mathrm{p_0 +2n\hbar k}$ (where $\mathrm{p_0}$ is the initial momentum of BEC, k is the wave number of light  and  n is the order of Bragg transition) through a  two-photon stimulated process \cite{PhysRevLett.82.871}. We perform the AI in Mach-Zehnder ($\mathrm{\pi/2-\pi-\pi/2}$) configuration \cite{PhysRevA.61.041602} where  $\mathrm{\pi/2}$ pulse acts as beam splitter  and $\mathrm{\pi}$ pulse is used as mirrors \cite{PhysRevLett.100.180405}. A brief experimental procedure is shown in Figure \ref{fig1} based on references \cite{PhysRevLett.117.203003, PhysRevA.84.033610, PhysRevLett.75.2638}. In order to enable atomic gravimetry, one can introduce two counter-propagating or co-propagating beams that create a moving optical lattice i.e. Bragg beams which can be realised with beams of  variable frequency components f and $\mathrm{f+\delta f}$. In the absence of any external force on atoms, the relative acceleration between the moving optical lattice and atoms remains uniform, the paths followed by the atoms during the interferometer will be identical, resulting in a zero phase contribution. Consequently, in the presence gravitational force on atoms, the overall phase shift will be proportional to the uniform acceleration generated by the interaction between light and atoms, as described in reference as \cite{PhysRevA.84.033610}: 
\begin{equation}
\mathrm{\Phi=n(\phi_1-2\phi_2+\phi_3)=2n \mathbf{k}.\mathbf{g}T^2}
\label{eq1}
\end{equation}
Here, the optical phases $\mathrm{\phi_1},\mathrm{\phi_2}$ and $\mathrm{\phi_3}$  represents the interactions of atoms with Bragg pulses, n corresponds to the Bragg order, and T signifies the interferometric time of the atom interferometer (AI). Scanning the phase of final  $\pi/2$ pulse , we observe oscillation in the population of both momentum order, and the  resulting signal  exhibits to:
\begin{equation}
\mathrm{P=N(1+C\cos(\Phi))/2}
\label{eq2}
\end{equation} where N is the population of atoms and C is the contrast of signal. The fundamental concept of measuring gravity is to balance the phase difference imparted to the atoms by gravitational acceleration and the Bragg AI pulses. As freely falling atoms experience a time-dependent Doppler shift with respect to the Bragg transition, the optical lattice is accelerated by adjusting the frequency difference in the lattice beam. The overall phase shift can be obtained by scanning the lattice acceleration around the local gravity which becomes as  $\mathrm{\Phi=n(2\mathbf{k}.\mathbf{g}T^2-2\pi\alpha T^2)}$ where $\alpha$ is the sweep rate (also considered as frequency chirp) of lattice. To determine the value of g, the value of  $\alpha$ (say, $\alpha_{0}$)  will balance the gravity and overall phase shift $\Phi$ will be zero, thus providing $\mathrm{\alpha_{0}=\frac{1}{\pi}(\mathbf{k}.\mathbf{g})}$. To determine the value of $\mathrm{\alpha_{0}}$, one has to observe the interferometric signal for at least three different interferogram time(T), and all those interferometric signal have a common minima at $\alpha_{0}$. 

\subsection{Noise Model of an Interferometer due to its optical path}
A typical or conventional atom interferometer using Bragg-based principles is constructed with a heterodyne optical setup \cite{hardman2016bec}. This configuration involves splitting a single beam into two separate paths and directing them through two acousto-optic modulators (AOMs) with frequencies  $\mathrm{f_{1}}$  and $\mathrm{f_{2}}$ respectively. The resulting heterodyne interferometric signal is detected and can be described by the equation:
\begin{equation}
\mathrm{S(t)= A \left[1+C\cos(2\pi f_{c}t+\phi_{n}(t)+\phi_{0})\right]}
\label{eq3}
\end{equation} 

where A represents the amplitude of the DC component,  C is the visibility or contrast, $\mathrm{f_{c}=f_{1}-f_{2}}$ denotes the heterodyne frequency, and $\mathrm{\phi_{0}}$ represents the average phase. The term $\mathrm{\phi_{n}(t)}$ represents the phase noise induced by acoustic, optical, and electronic components. To extract the phase noise, one utilizes the orthogonal demodulation technique. This technique involves mixing S(t) with a low-noise reference frequency using a lock-in amplifier. The lock-in amplifier mixes $\mathrm{S(t)}$ with $\mathrm{\sin(2\pi f_{c}t)}$ and $\mathrm{\cos(2 \pi f_{c}t)}$, followed by a low-pass filter to eliminate undesirable higher frequencies. Consequently, Equation \ref{eq3} can be modified as proposed in \cite{8379416} to include the respective noise model:
\begin{equation}
\mathrm{S(t)= A[1+n_{m}(t)] [1+C\cos(2\pi f_{c}t+n_{p}(t)+\phi_{0})]+n_{a}(t)}
\label{eq4}
\end{equation} 

where $\mathrm{\phi_{n}(t)}$ is now represented as $\mathrm{n_{p}(t)}$ to account for phase noise originating from the laser source, AOM driver, and seismic vibrations. Additionally, $\mathrm{n_{a}(t)}$ accounts for additive noise, including amplified spontaneous emission noise, quantization noise, and circuit noise. Finally, $\mathrm{n_{m}(t)}$ represents multiplicative noise, which includes the amplitude noise of the optical pulse caused by the AOM driver and the relative intensity noise (RIN) of the laser. Our particular focus was on the noise induced by the AOM and optical elements resulting from acoustic vibrations.

\subsection{Sensitivity Function}
The sensitivity function gives the information regarding the atom interferometer phase shift $\mathrm{\Phi}$ due to infinitesimal laser phase shift $\mathrm{\delta\phi}$ in the Bragg pulses and thus the population measured at the interferometric outputs. The sensitivity function $\mathrm{g_{\phi}(t)}$ is defined as \cite{4444746}:

\begin{equation}
\mathrm{g_{\phi}(t) = \lim_{\delta\phi\to 0 } \frac{\delta\Phi(\delta\phi,t)}{\delta\phi}=\frac{2}{\sin\Phi}\lim_{\delta\phi\to 0 }\frac{\delta P(\Phi,\delta\phi,t)}{\delta\phi}}
\label{eq5}
\end{equation}

The above relation can be replicated for an interferometer both for beam splitters and mirrors as follows: 
\begin{equation}
\mathrm{g_{\phi}(t) = \sin\bigg(\int_{t_{0}}^{t}\Omega_{R}(t')dt'\bigg)}
\label{eq6}
\end{equation}
\begin{figure}[ht]
	\includegraphics[width=\linewidth]{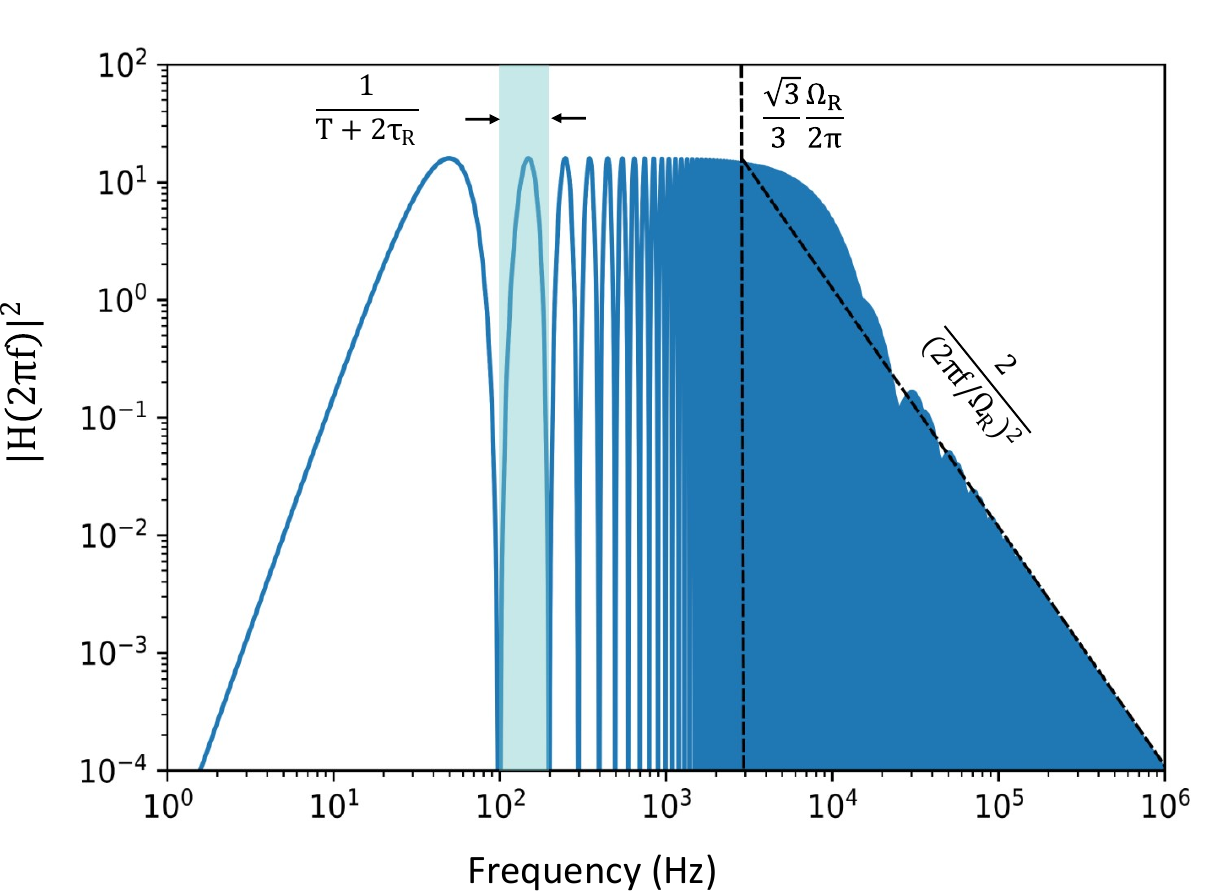} 
	\captionsetup{%
		justification=centerlast,
		labelfont =bf,
		singlelinecheck=false
	}
	\caption{Transfer function $\mathrm{|H(2\pi f)|^{2}}$ for a Mach-Zehnder sequence $\mathrm{\pi/2-\pi-\pi/2}$ with a Rabi frequency of $\mathrm{2\pi\times}$5 kHz for an interferometeric time of 10 ms. $\mathrm{\frac{n}{T+2\tau_{R}}}$ represents the frequencies where the sensitivity diminishes, where n is an integer and  $\mathrm{\frac{\sqrt{3}\Omega_{R}}{6\pi}}$ corresponds to the cutoff frequency for a finite duartion of Bragg pulses. }
	\label{fig4}
\end{figure}
where $\mathrm{\Omega_{R}(t)}$ is the Rabi frequency during the light-atom interaction in the interferometric sequence.
The complete value of $\mathrm{g_{\phi}(t)}$ depends on the scheme one uses to perform the interferometer. For a Mach-Zehnder interferometer with three consecutive pulses of $\pi/2$ and $\pi$, we consider the time origin at the middle of the second Bragg pulse. Thus the sensitivity function can thus be read as for one half of the sequence where $\mathrm{\tau_{R}}$ is the duration of the bragg pulse seperated by interferometeric time T:
\begin{equation}
\mathrm{g_{\phi}(t)}=\begin{cases}\mathrm{\sin(\Omega_{R}t)},& \mathrm{0 < t< \tau_{R}}\\
1, & \mathrm{\tau_{R} < t< T+\tau_{R}}\\
\mathrm{-\sin\left(\Omega_{R}(T-t)\right)},& \mathrm{T+\tau_{R} < t< T + 2\tau_{R}}\end{cases}
\label{eq7}
\end{equation}
Thus, the sensitivity function is used to determine the interferometric phase shift $\Phi$  for an arbitrary Bragg phase noise $\mathrm{\phi(t)}$ as:
\begin{equation}
\mathrm{\Phi = \int_{-\infty}^{+\infty}g_{\phi}(t)d\phi(t)=\int_{-\infty}^{+\infty}g_{\phi}(t)\frac{d\phi(t)}{dt}dt}
\label{eq8}
\end{equation}

We establish the interferometer's transfer function in the Fourier domain $\mathrm{H(\omega)=H(2\pi f)=\omega G(\omega)}$ where $\mathrm{G(\omega)=\int_{-\infty}^{+\infty}e^{-i\omega t}g_{\Phi}(t)dt}$ is the fourier transform of the sensitivity function defined as: 
\begin{equation}
\begin{split}
\mathrm{G(\omega)} & \mathrm{=\frac{4i\Omega_{R}}{\omega^2-\Omega_{R}^2}\sin\left\{\frac{w(T+2\tau_{R})}{2}\right\}} \\
&\mathrm{\times \left[\cos\left\{\frac{w(T+2\tau_{R})}{2}\right\}+\frac{\Omega_{R}}{\omega}\sin\left(\frac{\omega T}{2}\right)\right]}
\label{eq9}
\end{split}
\end{equation}

To assess how the laser phase noise affects the interferometer sensitivity, we thus defined the rms standard deviation of the phase noise in the interferometer as:

\begin{equation}
\mathrm{(\sigma_{\Phi})^2=\int_{0}^{+\infty}|H(\omega)|^2 S_{\Phi}(\omega)d\omega}
\label{eq10}
\end{equation}

where, $\mathrm{S_{\phi}(\omega)}$ represents the power spectral density of phase of the Bragg phase.

Hence, the transfer function was plotted in Figure \ref{fig4} to analyze its behavior in relation to frequency. The transfer function exhibits oscillatory patterns, providing insights into the diminishing amplitude of repetitive regions at frequencies given by $f=n/(T+2\tau_{R})$, where $n$ is an integer. This behavior indicates that the interferometer functions acts as a low-pass filter, with a cutoff frequency defined as $f=\sqrt{3}\Omega_{R}/6\pi$. As the frequency increases, the transfer function exhibits a trend proportional to $2/(2\pi f/\Omega_{R})^2$, leading to a decrease in interferometric sensitivity.


\section{Experimental Details}

\begin{figure*}[!tbp]
	
	\subfloat[]{\includegraphics[width=0.4\textwidth]{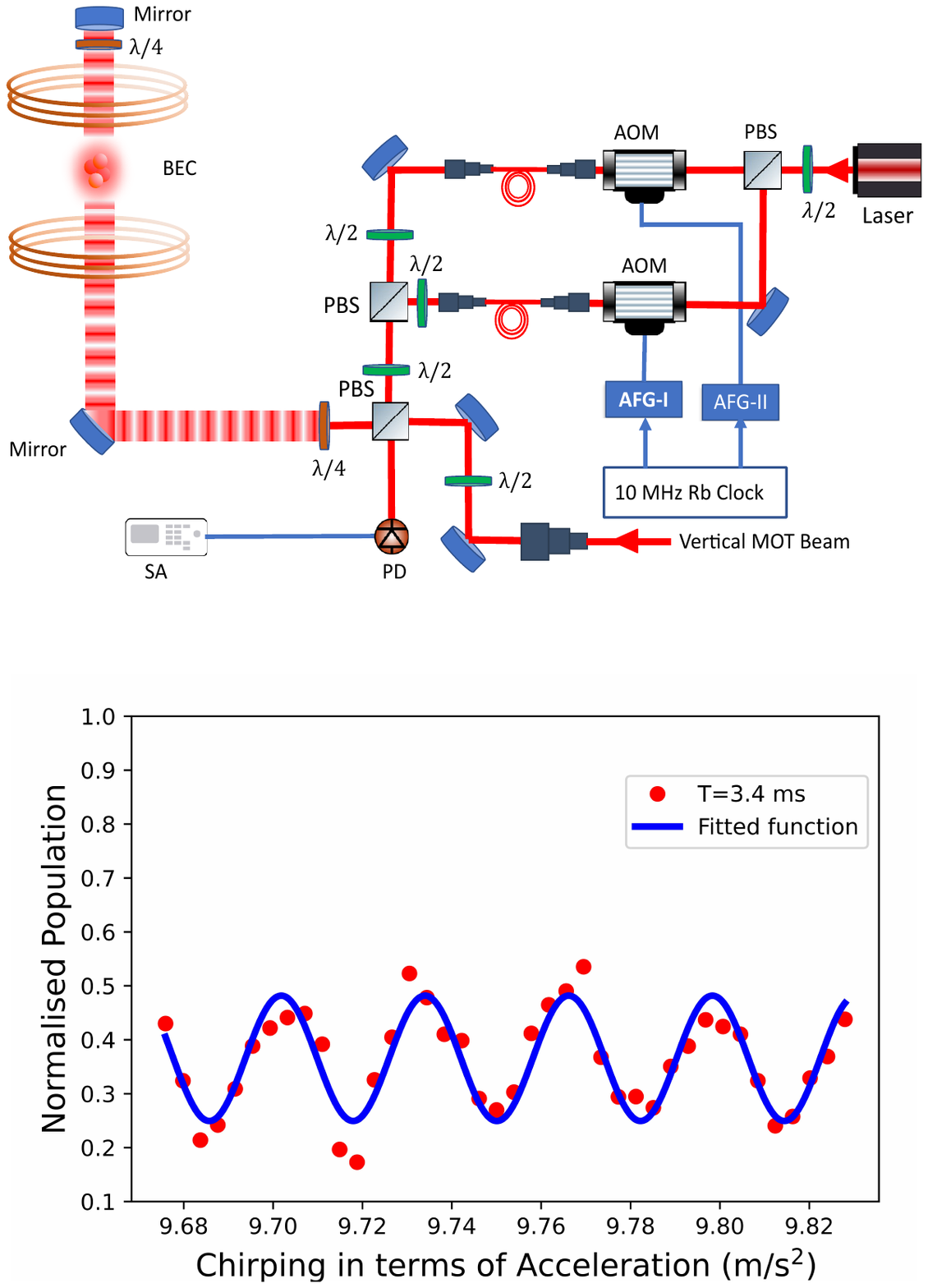}\label{fig:f1}}
	\qquad
	\subfloat[]{\includegraphics[width=0.4\textwidth]{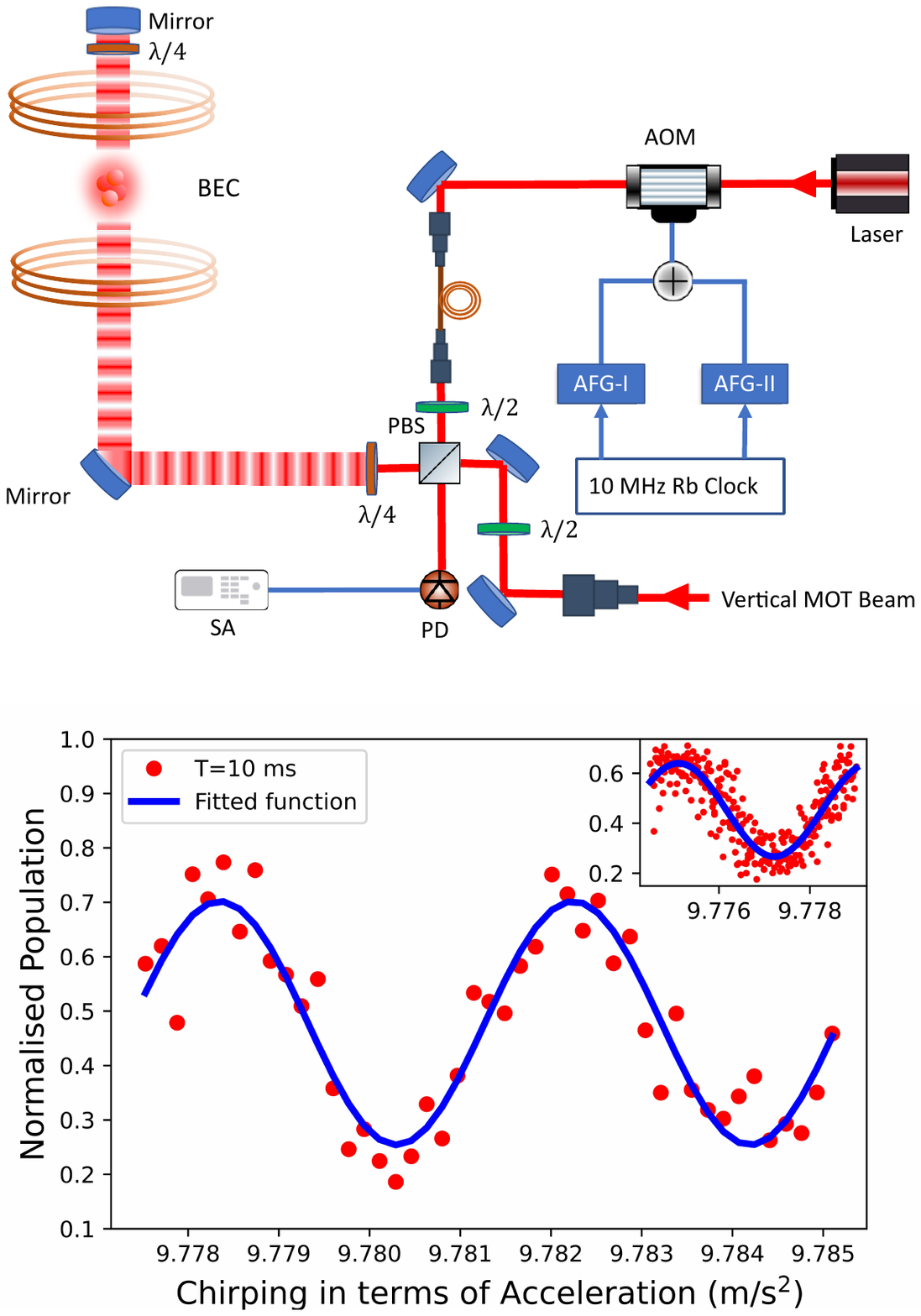}\label{fig:f2}}
	\captionsetup{%
		justification=centerfirst,
		labelfont =bf,
		singlelinecheck=false
	}
	\caption{ {Schematic diagram of generation of Bragg beams and interferometric signal for two different configurations}. \textbf{(a)} Top figure depicts the  block diagram of generation of Bragg pulses using two different AOMs driven by AFGs which are phase locked using Rubidium atomic clock. The laser is locked to the Rubidium transition line which is not shown in the diagram. PD: photodiode, SA: spectrum analyser, AFG: arbitary function generator. Bottom figure shows the population oscillation of first order momentum state versus sweep rate in terms of acceleration for interferometer time T = 3.4 ms. \textbf{(b)} Top figure depicts the block diagram of generation of Bragg pulses using single AOM driven by AFG which are phase locked. Bottom figure which is adapted from our previous work \cite{dutta2023decade} shows the population oscillation of first order momentum state versus sweep rate in terms of acceleration  for interferometer time T = 10 ms for single AOM configuration. Inset: Precise scanning of sweep rate along with the interferometric signal. }
	\label{fig2}
\end{figure*}

 We create a BEC of $\mathrm{^{87}Rb}$ consisting of $5\times10^4$ atoms every 15 seconds in a dipole trap. The temperature of the residual thermal component of the BEC is measured to be less than 200 nK. The experimental setup is almost the same as in this reference \cite{PhysRevE.106.034207}. After turning off the dipole trap, we provide  2 ms time of flight to reduce BEC's mean-field effect on AI. The laser used for realizing the optical lattice is locked to the $\mathrm{5^{2}S_{1/2}, F = 2 \longrightarrow 5^{2}P_{3/2}, F'=2}$, D2 transition at 780 nm. Since the BEC is prepared in the $\mathrm{5^{2}S_{1/2}, F =1, m_{F} = -1}$ state, the laser is 6.8 GHz red-detuned from the atoms’ accessible transition to suppress spontaneous emission. Now, for operation as a gravimeter, we have introduced two different methods where a vertical, linearly polarized light beam travelling through two differnt Acousto-Optic Modulator (AOM) (ATM-801A2) or a single Acousto-Optic Modulator (AOM), which is driven by Arbitrary Function Generators (AFG) (AFG3032C)  phase-locked by Rubidium frequency standard (FS725) as shown in the Figure \ref{fig2} (a) and \ref{fig2} (b). By employing the AFG to drive the AOM, we gain precise control over various parameters such as frequency, sweep rate in frequency, and the phase of the lattice beam. To drive the Bragg transition in $^{87}Rb$, the frequency difference in lattice beam should be $\mathrm{\delta f=4n\omega_{R}}$, where n is the order of Bragg transition and $\mathrm{\omega_{R}}$ is recoil frequency, and frequency difference is about 15 kHz in our experimental setup for first-order Bragg transition. But when the atoms are freely falling under gravity, the Bragg transition condition gets modified because atoms feel a time-dependent Doppler shift $\mathrm{\delta_{d}(t)= 2\pi \alpha_{0}t}$, where $\mathrm{\alpha_{0} =(\frac{1}{\pi})(\mathbf{k}.\mathbf{g})}$ is a frequency chirp. The resonance condition for the Bragg transition is then transformed in laboratory frame as $\mathrm{\delta f = 4n\omega_{R} + 2(\mathbf{k}.\mathbf{g})t}$ \cite{PhysRevA.84.033610}. To compensate for this Doppler shift, we apply a 25.078 MHz/s sweep rate (determined by the approximate theoretical value of g in Pune) in one of the lattice beams to keep the atoms on resonance for the Bragg transition. For generating the AI pulses $\mathrm{(\pi/2-\pi-\pi/2)}$ in the Mach-Zehnder configuration, we have used square pulses with a pulse duration of 50 $\mathrm{\mu s}$ for $\pi/2$ pulses and 100 $\mathrm{\mu s}$ for $\mathrm{\pi}$ pulse to drive the first-order Bragg transition, and a time sequence has shown in Figure \ref{fig1} with a $1/e^2$ beam diameter of about 2.5 mm. The typical power in each beam near the interrogation site is about 1 mW. The effective Rabi frequency corresponding to this beam intensity is calculated as  $\mathrm{\Omega_{eff}=\frac{\Omega_{1}\Omega_{2}}{2\Delta}\simeq 2\pi\times}$5  kHz where $\mathrm{\Omega_{1}}$ and $\mathrm{\Omega_{2}}$ represent the resonant Rabi frequencies of two Bragg beams,and $\Delta$ denotes the detuning of the beams from the optical transition.
 
Figure \ref{fig2} represents the interference fringes obtained from two different configurations. Figure \ref{fig2} (a) interference signal is the first configuration involved the utilization of two co-propagating laser beams diffracted by AOMs operating at 80 MHz and 80.015 MHz to generate an optical lattice as illustrated in reference \cite{10.1063/1.5001963} without any active feedback locking technique. The interference pattern exhibited an interferometric oscillation with low contrast, resulting in a transfer efficiency of approximately $\sim25\%$ for an interferometric time of T = 3.4 ms. This reduction in contrast  or visibility was attributed to phase noise induced by acoustic and sub-acoustic vibrations coupled to the atomic system.

In the second configuration, a single AOM was employed, and dual frequencies of 80 MHz and 80.015 MHz were introduced from two phase-locked AFGs, as illustrated in Figure \ref{fig2} (b). Notably, this configuration demonstrated an improvement in contrast or visibility compared to the previous method. Thus the graph depicts the population oscillation of $\mathrm{p = 2\hbar k}$ as a function of sweep rate in terms of acceleration, exhibiting a transfer efficiency of approximately $\sim50\%$ for an interferometric time of T = 10 ms. The inset figure provides a detailed scan in relation to the sweep rate.


\section{Results and discussion}
\subsection{Phase Noise for different configuration}
With a interferometric cycle time of approximately 2T = 20 ms our interferometer is sensitive to fast noise contribution $\Phi_{noise}$ down to 50 Hz. It is important to note that due to geometric limitations in our system, we were unable to extend the interferometric experiments beyond a 40 ms time of flight. The principal source of noise in the measurement is the phase noise of the laser which interacts with the atoms. This laser phase originates from laser source and from the vibrations that shift the phase fronts of the two co-propogating laser beams. Since in our case we use Bragg beams which is originated from a single laser source due to which the contribution due to it is less compared to acoustic vibrations.

\begin{figure}[ht]
	\centering
	\includegraphics[width=\linewidth]{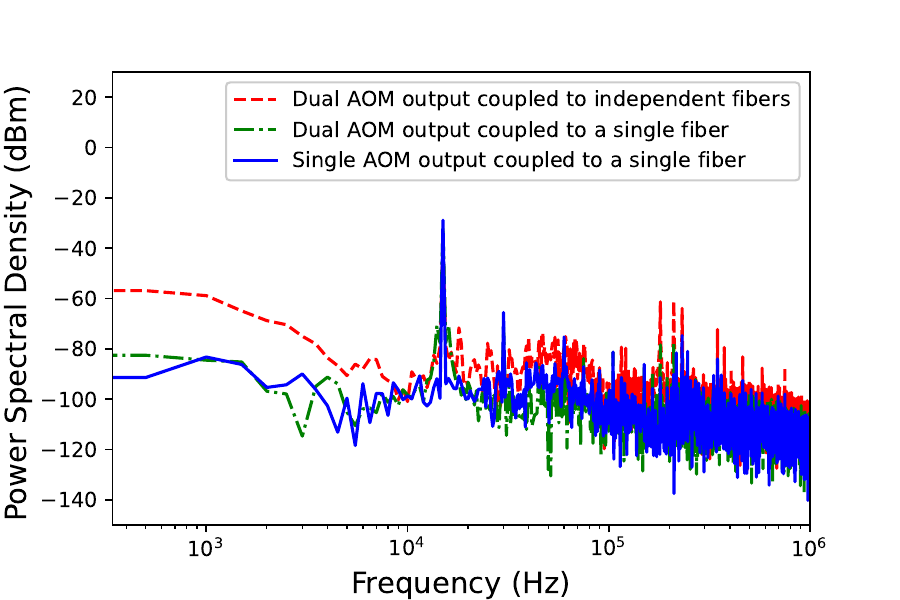} 
	\captionsetup{%
		justification=centerlast,
		labelfont =bf,
		singlelinecheck=off
	}
	\caption{The power spectral density of the Bragg beams is examined for various configurations. In this analysis, the dashed red line represents the configuration involving two co-propagating laser beams diffracted from separate AOMs as shown in  Figure \ref{fig2} (a), the dot-dashed green line represents the configuration of Bragg beams diffracted from seperate AOMs and coupled into a single fiber to reduce the differential phase noise and the solid blue line corresponds to the configuration utilizing a single AOM as shown in the Figure \ref{fig2}(b), with a frequency resolution of 500 Hz. }
	\label{fig5}
\end{figure}

To evaluate the contribution of different configuration we set the frequency difference between the lattice beam at 15 kHz, the first order Bragg resonance for $\mathrm{^{87}Rb}$. The beat signal is logged at a sampling rate of 2.5 Giga samples per second for 2 ms with a record length of 5 million points. A fourier transform of the logged data is coverted into the power spectrum shown in Figure \ref{fig5}. The dashed red line shows the power spectrum of  configuration, as depicted in Figure \ref{fig2} (a), exhibits power spectral density above -80 dBm for frequencies below 10 kHz well above the single AOM configuration as shown in Figure \ref{fig2} (b) which is depicted in solid blue line.  The power spectral density analysis was conducted with a frequency resolution of 500 Hz. We also compared the beat signal generated by two AOMs and coupled to a single fiber to reduce the differential phase noise which is represented as the dot-dashed green line.

Thus, the initial experimental setup, depicted in Figure \ref{fig2} (a), resulted in lower interferometric signal due to the presence of phase noise induced by vibrations, which resulted in a constraint on the interferometric time due to loss of contrast. And, remarkably, Figure \ref{fig2} (b) demonstrates a notable enhancement in both contrast and interferometric signal when we transitioned to the configuration utilizing a single AOM.
\begin{figure}[ht]
	\centering
	\includegraphics[width=\linewidth]{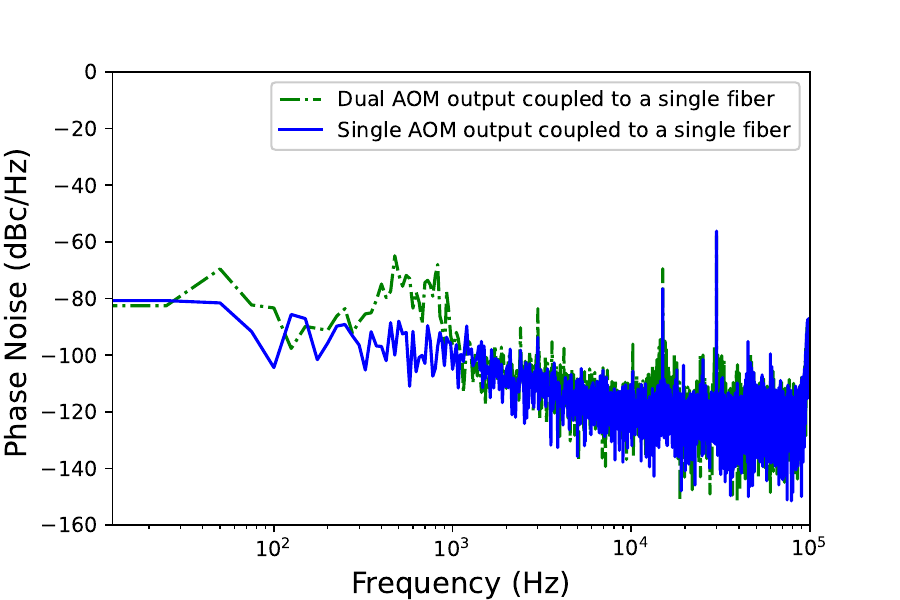} 
	\captionsetup{%
		justification=centerlast,
		labelfont = bf,
		singlelinecheck=off
	}
	\caption{Phase Noise spectrum of te Bragg beams for single AOM configuration and dual AOM configuration coupled to a single fiber with 25 Hz frequency resolution.  }
	\label{fig6}
\end{figure}
\par

We had also presented the phase noise spectrum for single AOM configuration and double AOM output coupled to a single fiber with a frequency resolution of 25 Hz. Notably, the single AOM configuration, represented by the blue line in Figure \ref{fig6}, achieves a significant suppression of residual phase noise by two orders of magnitude around 800 Hz. The calculated integrated phase noise in the Bragg pulse within the frequency range upto 10 kHz from Equation \ref{eq10} for single AOM configuration, measures 10 mrad/shot, while for the dual AOM configuration coupled to a single fiber  amounts to 47 mrad/shot.

\begin{figure}[ht]
	\centering
	\includegraphics[width=\linewidth]{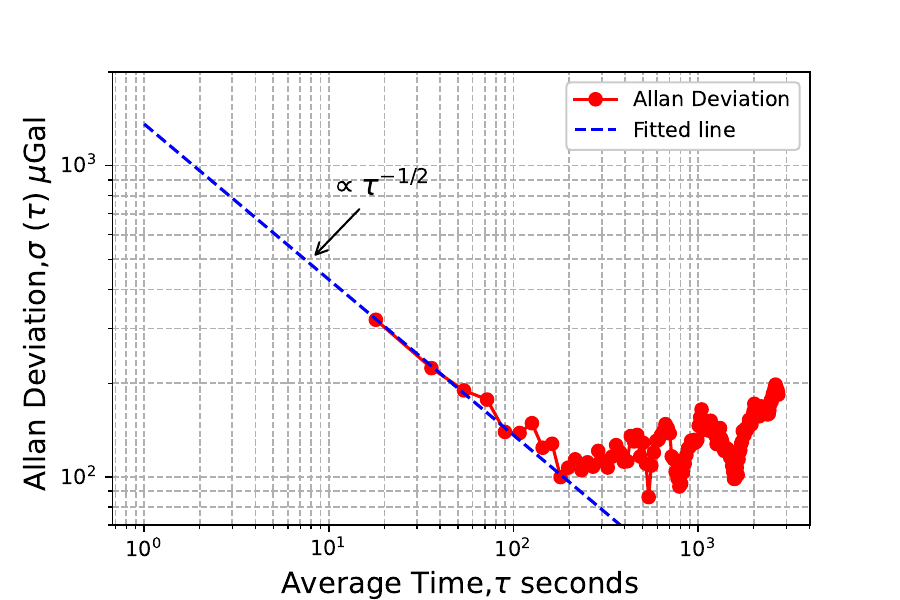} 
	\captionsetup{%
		justification=centerlast,
		labelfont = bf,
		singlelinecheck=off
	}
	\caption{Allan deviation of the gravity measurements  for interferometric time  $\mathrm{T =10}$ ms. The dashed line corresponds to a short-term sensitivity of 1360  $\mathrm{\mu Gal/\sqrt{Hz}}$ for 1 second.}
	\label{fig7}
\end{figure}
\subsection{Stability of the gravity measurement}
To justify the resolution of the determined gravity value we performed the integration of the stability of the experiment. The AI interferometer is operated for about 2 hours with a pulse seperation time of $\mathrm{T=10}$ ms for cycle time of $\mathrm{T_{cycle} =17.98}$ s.  Figure \ref{fig7}  depicts the time series data in the form of Allan deviation of the population of atom in first order. 
The phase noise for the atomic interferometer with the better configuration is calculated using an Allan deviation. This which is derived as the square root of the Allan variance,
\begin{equation}
\begin{split}
\sigma_{y}^{2}(\tau) = \frac{1}{2(M-1)}\sum_{i=1}^{M-1}(y_{i+1}-y_{i})^2
\label{eq17}
\end{split}
\end{equation}
for a collections of M mean data points $y_{i}$, acquired at average interval $\tau$. For this anlaysis, the data points corresponds to the measured acceleration due to gravity and the averaging time is expressed in units of runs that corresponds to the 17.9 s duty cycle of the experiment. Thus Allan deviation is a standard tool for assessing the temporal characteristics of noise in precision measurements.Hence, the Allan deviation serves as a common tool for evaluating the temporal properties of noise in precise measurements.

The measured short-term sensitivity of ultracold atom interferometer is estimated to be 1360 $ \mathrm{\mu Gal/\sqrt{Hz}}$ which is extrapolated to  1 second according to the white noise behaviour as it scales as $\tau^{-1/2}$ where $\tau$ is the average time of operation. The allan deviation decreases down to 99.7  $\mathrm{\mu Gal/\sqrt{Hz}}$ for an integration time of 200 seconds.

For the given interferometric time of T=10 ms, we have estimated the intrinsic sensitivity limit of the interferometer as \cite{abend2017atom}:
\begin{equation}
(\mathrm{\Delta g/g)_{limit} = \sigma_{qpn}.\sigma_{g} = \frac{1}{C\sqrt{N}gk_{eff}T^2}}
\label{eq18}
\end{equation}

where $\mathrm{(\Delta g/g)_{limit}}$ is the sensitivity of the system, $\mathrm{\sigma_{qpn}}$ is the quantum projection noise, $\mathrm{\sigma_{g}}$ is the scaling factor of the interferometer to changes in g and $\mathrm{k_{eff}=2k}$. The calculated intrinsic sensitivity limit for $\mathrm{T=10}$ ms is obtained to be around $56.7\times10^{-8}$ for 50$\%$ contrast with $5\times10^4$ atoms.

\section{Conclusion}

In this study, we conducted a comparative analysis of two different techniques employed in atom interferometry (AI) for measuring local gravitational acceleration. Our approach involved utilizing a Mach-Zehnder matter wave interferometer, where Bragg diffraction of $^{87}$Rb atoms in the Bose-Einstein condensate (BEC) state was employed. By implementing the one AOM configuration instead of the conventional method, we successfully reduced the phase noise in our system. As a result, we were able to extract the fringe visibility or contrast for the atom interferometer. The sensitivity of our system was determined to be $\mathrm{99.7}$ $\mathrm{\mu Gal/\sqrt{Hz}}$, at an integration time of  200 seconds interval , with an interferometric time of 10 ms. Furthermore, we demonstrated that the conventional method, which involves splitting the beam and passing it through two AOMs coupled into a single fiber for noise cancellation, exhibited higher integrated phase noise compared to the single AOM configuration.
In summary, our findings highlight the efficacy of employing a one AOM configuration to reduce phase noise, improve fringe visibility, and enhance the overall performance of the atom interferometer system for measuring local gravitational acceleration.
\phantomsection
\section*{Acknowledgments} 

\addcontentsline{toc}{section}{Acknowledgments} 

The authors would like to thank the Department of Science and Technology, Govt. of India and the Council for Scientific and Industrial Research for research funding. PD and SSM would like to thank Council for Scientific and Industrial Research research fellowships. KB and KP would like to thank IISER Pune for institutional fellowships. The authors would like to express their gratitude  and acknowledge Sumit Sarkar and Jay Mangaonkar for their fruitful and valuable discussions. The authors would also like to acknowledge the funding support from I-HUB Quantum Technology Foundation through the National Mission on Interdisciplinary Cyber-Physical Systems (NM-ICPS) of the Department of Science and Technology, Govt. of India.




\section*{References}


\end{document}